\documentclass{pasa}%

\title[Young Massive Binary DN~Cas]{Study of Eclipsing Binary and Multiple Systems in OB Associations IV: Cas~OB6 Member DN~Cas}
\author[Bak{\i}s et al.]{V. Bak{\i}\c{s}$^1$\thanks{volkanbakis@akdeniz.edu.tr}, H. Bak{\i}\c{s}$^1$, S. Bilir$^2$, Z. Eker$^1$\\
\affil{$^1$Department of Space Sciences and Technologies, Faculty of Science, Akdeniz University, 07058, Antalya, Turkey} %
\affil{$^2$Department of Astronomy and Space Sciences, Faculty of Science, Istanbul University, 34119, Beyaz\i t, Istanbul, Turkey}} %
\jid{PASA}
\doi{10.1017/pas.\the\year.xxx}
\jyear{\the\year}
\usepackage{natbib}
\begin{document}%

\begin{abstract}
An early-type, massive, short-period ($P_{orb}=2^{\rm d}.310951$) eclipsing spectroscopic binary DN~Cas has been re-visited with new spectral and photometric data. The masses and radii of the components have been obtained as $M_1=19.04\pm0.07 M_\odot$, $M_2=13.73\pm 0.05M_\odot$ and $R_1=7.22\pm0.06 R_\odot$, $R_2=5.79\pm0.06R_\odot$, respectively. Both components present synchronous rotation ($V_{rot1}=160\,$km s$^{-1}$, $V_{rot2}=130\,$km s$^{-1}$) with their orbit. Orbital period analysis yielded a physically bound additional component in the system with a minimum mass of $M_3=0.88 M_\odot$ orbiting in an eccentric orbit ($e=0.37\pm0.2$) with an orbital period of $P_{12}=42\pm9$ yrs. High precision absolute parameters of the system allowed us to derive a distance to DN~Cas as 1.7$\pm$0.2 kpc which locates the system within the borders of the Cas~OB6 association ($d=1.8$ kpc; Mel'nik \& Dambis, 2009). The space velocities and the age of DN~Cas are in agreement with those of Cas~OB6. The age of DN~Cas ($\tau=$3-5 Myr) is found to be 1-2 Myr older than the embedded clusters (IC~1795, IC~1805 and IC~1848) in the Cas~OB6 association, 
which implies a sequential star 
formation in the association.
\end{abstract}

\begin{keywords}
Galaxy: open cluster and associations: individual: Cas OB6 -- stars:
early-type -- stars: individual: DN~Cas
\end{keywords}
\maketitle%
\section{Introduction}

OB associations host the gravitationally unbound youngest and massive stars in the galaxy, therefore, stellar parameters of the members of OB associations help understanding their structure and evolution. The most powerful technique to obtain age, chemical properties and a reliable distance is doubtlessly the use of eclipsing binaries, which are the members of the association. Luckily, the relative number of eclipsing binary and multiple (EBM) systems in OB associations is high (e.g. Mason et al., 1998). Within the scope of investigating EBM systems in OB associations with modern analysis techniques (i.e. Bak{\i}\c{s} et al., 2015), an early-type massive binary DN~Cas in the direction of Cas~OB6 complex is selected for a more detailed study than available in the literature today.

Located in the front edge of the Perseus spiral arm, the Cas~OB6 association is one of the most outstanding associations in the Cas~OB complex (see Fig.~1). Its distance and mean radial velocity were determined by Mel'nik \& Dambis (2009) as 1.8 (kpc) and --42.7$\pm$8.2 km s$^{-1}$, respectively. The association hosts a number of HII regions (IC 1805, IC 1848, S196 and S201) and several radio sources W5 (IC 1848), W4 (surrounding IC 1805) and the W3 region (IC 1795). These regions are known with massive stellar content and active stellar formation which goes back in time to 3-5 Myrs for IC 1795 (Bik et al. 2012), 1-3 Myrs for IC~1805 (Strai\v{z}ys et al., 2013) and 3-5 Myrs for IC~1848 (Lim et al., 2014).

DN~Cas is located within the borders of Cas~OB6. Its photometric variation was discovered by Hoffmeister (1947) and has been observed photometrically by Frazier \& Hall (1974) for the first time. Due to variability of the selected comparison star, Frazier \& Hall (1974) was not able to solve the multiband light curves. Davidge (1980) performed $UBV$ photometry of the system without the light curve solution. Zakirov (2001) who investigated the relationship of DN~Cas with Cas OB6 made estimations on the absolute parameters of the companions using his own photometric data in $UBVRI$-bands. However, due to absence of the spectral information, the estimated parameters were not consistent with the spectral types of the components.

The most recent study on DN~Cas is the radial velocity survey of the 14 association members of Cas~OB6 (Hillwig et al. 2006). They reported a binarity of 50 percent in their sample and obtained the spectroscopic orbit of DN~Cas for the first time. Using the orbital parameters from the analysis of incomplete monochromatic light curve of DN~Cas, stellar parameters of the components of DN~Cas were estimated. However, necessity of multiband photometric observations of the system was announced probably due to the need to clarify the inconsistency of the mass and spectral type of the components they derived.

\begin{figure*}
	\begin{center}
		\begin{tabular}{c}
			\resizebox{120mm}{!}{\includegraphics{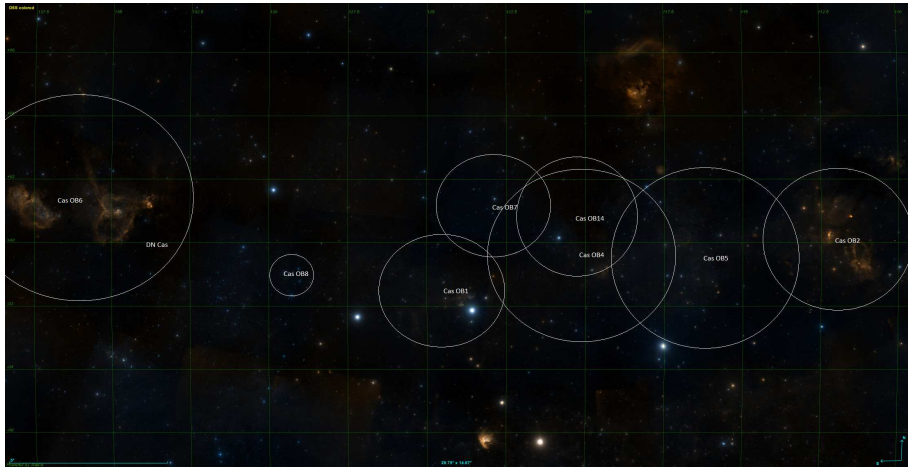}} \\
		\end{tabular}
		\caption{The distribution of the Cas~OB associations towards galactic coordinates $l=120^\circ.3$ and $b=-0^\circ.9$, created with the \textit{Aladin Sky Atlas} (Bonnarel et al., 2000). The borders of individual associations were determined using the catalogue data of Mel'nik \& Dambis (2009) and Tetzlaff et al. (2010).}\label{fig1}
	\end{center}
\end{figure*}

\section{Observations}

\subsection{Photometry}

There are three sources of photometric data of DN~Cas used in this research: the first set of photometric data comes from the observations of Zakirov (2001) which is not available in the literature. The observations of Zakirov (2001) were used in this paper for only orbital period study (\S3). The second set of photometric data was collected from the observations which were performed in \c{C}anakkale Onsekiz Mart University Ulup{\i}nar Observatory in 2009 with 30-cm and 40-cm Cassegrain-Schmidt type telescopes equipped with 1k$\times$1k CCDs and Bessell $UBVR_cI_c$ filters. A total of 4956 images have been collected in 7 nights. The data obtained in this campaign was not evenly distributed in phase, therefore, they were used for only orbital period analysis with a new time of minimum (HJD~2455128.3279$\pm$0.0005) (see \S3).

The third set of photometric data was collected during the observing campaign in Akdeniz University Campus Telescope (AUCT) in 2012 using a 25-cm Ritchey-Chretien type telescope equipped with a back-illuminated 1k$\times$1k CCD camera and Bessell $UBVR_cI_c$ filters. A total of 3840 images has been collected in 9 nights. The number of data and its distribution are enough to analyze the multi-band light curves of DN~Cas. The mean standard deviation of the observations are 0.033, 0.011, 0.009, 0.007 and 0.010 in $UBVR_c$ and $I_c$ filters, respectively. The comparison star HD~14558 was selected for differential photometry and the check star was SAO~12245. The reduction of the photometric data is standard for CCD photometry; bias and dark subtraction and flat field correction. The photometry of the reduced data were performed by means of aperture photometry where the star aperture was selected to be 3 times the FWHM of the stellar profile. A new time of primary minimum has been extracted from the AUCT 
observations (HJD~2456287.2716$\pm$0.0003).

\subsection{Spectroscopy}

The spectroscopic observations has been obtained in 2013 with the intermediate resolution (R$\sim$5100) TUG Faint Object Spectrograph and Camera (TFOSC) attached to 1.5-m RTT150 telescope of T\"UB\.ITAK National Observatory\footnote{http://www.tug.tubitak.gov.tr/}, Turkey. TFOSC provides a wavelength coverage from $\lambda=$330 nm to 1200 nm in 11 spectral orders. A total of 20 spectra have been obtained with TFOSC. The reduction of TFOSC spectra has been made in Image Reduction and Analysis Facility ({\sc IRAF}\footnote{IRAF is distributed by the National Optical Astronomy Observatories.}) software. The journal of spectroscopic observations is given in Table~1 and two spectral orders obtained at three different epochs are shown in Fig.~2. The spectral data obtained in this study were used as complementary to the existing radial velocity data of Hillwig et al. (2006).

The spectral lines of the components of DN~Cas are well separated (see Fig.~2) which allowed us to measure the spectral lines of the individual components reliably. The radial velocities of DN~Cas were measured by means of fitting Gaussian to the line centers using the {\sc splot} task in {\sc iraf}. The uncertainties of the radial velocity measurements have been determined as the standard deviation of three consecutive measurements which is on the order of 2-3 km s$^{-1}$. However, we estimate overall radial velocity uncertainty in the range 5-9 km s$^{-1}$ due to the low resolution of the spectrograph. The measured radial velocities are given in Table~1.

\begin{figure*}
	\begin{center}
		\begin{tabular}{c}
			\hspace{1cm} \resizebox{80mm}{!}{\includegraphics{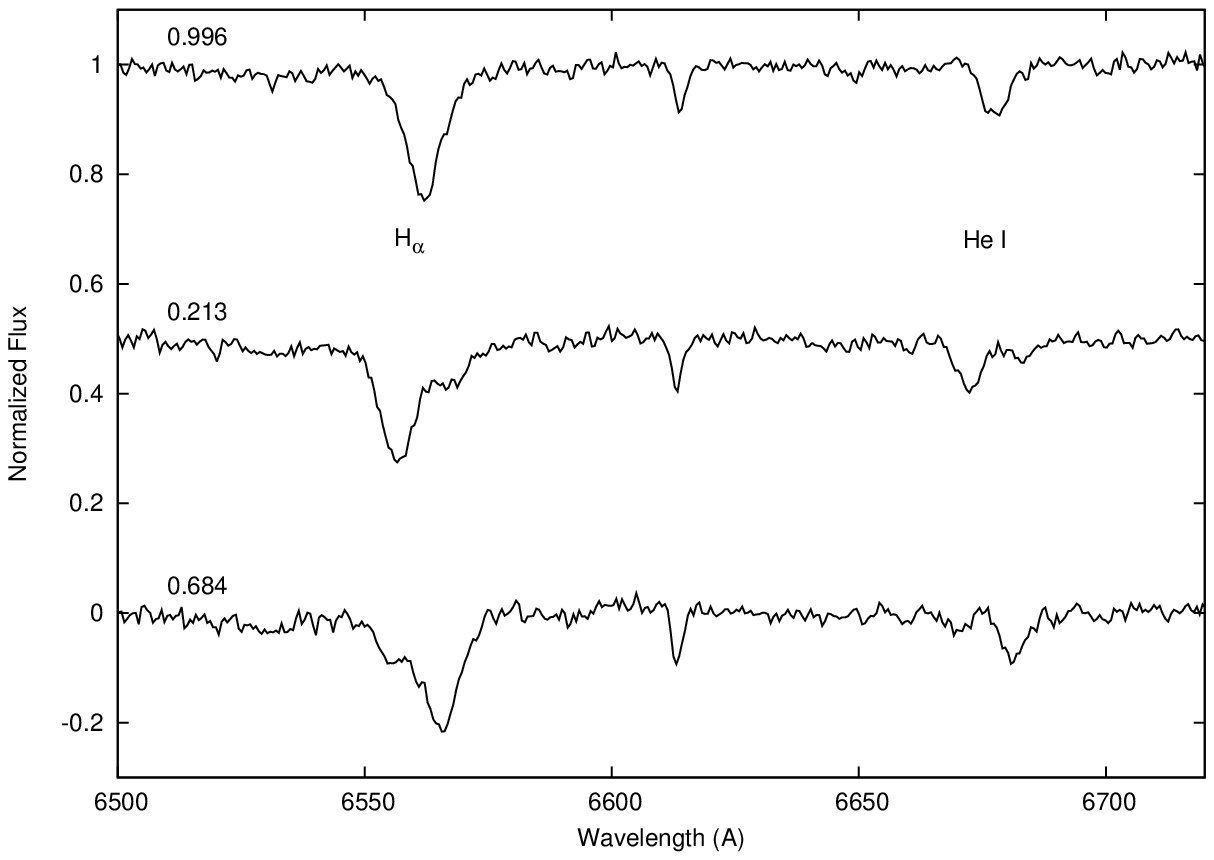}} \\
			\hspace{1cm} \resizebox{80mm}{!}{\includegraphics{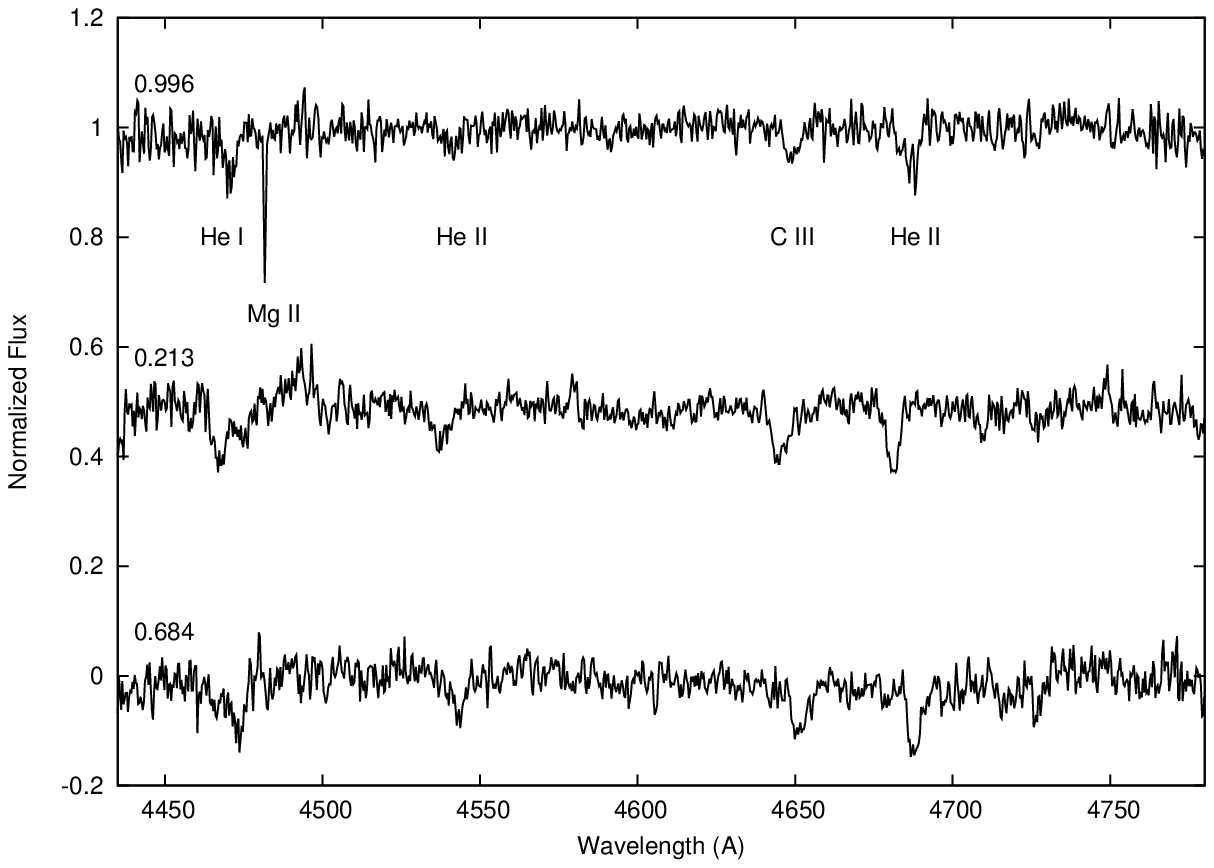}} \\
			\vspace{0.5cm}
		\end{tabular}
		\caption{Two spectral orders obtained at three different orbital phases.}\label{fig2}
	\end{center}
\end{figure*}

\begin{table*}
	\small
	\begin{center}
		\caption{Journal of spectroscopic observations and radial velocities (RV) of the components. \label{table1}}
		\begin{tabular}{lcccccc}
			\hline
			Number & Date         & Orbital Phase & Exposure Time & S/N & Pri RV & Sec RV \\
			& HJD-2400000  &               &  (s)          &     & (km s$^{-1}$) &  (km s$^{-1}$) \\
			\hline
			1	& 56303.4372& 0.997	&1200&110 & --66 & -\\
			2	& 56306.2212& 0.202	&1200&200 & --289 & 228\\
			3	& 56306.2386& 0.209	&1200&160 & --302 & 182\\
			4	& 56648.2582& 0.209	& 300&110 & --277 & 196\\
			5	& 56648.2642& 0.211	& 300&110 & --276 & 193\\
			6	& 56648.2717& 0.214	& 300&110 & --278 & 185\\
			7	& 56649.2205& 0.625	& 300&160 & 70 & --267 \\
			8	& 56649.2307& 0.629	& 300&140 & 51 & --306 \\
			9	& 56649.2403& 0.633	& 300&140 & 61 & --297 \\
			10	& 56649.2500& 0.637	& 300&130 & 61 & --311 \\
			11	& 56649.2588& 0.641	& 300&130 & 68 & --343 \\
			12	& 56649.2681& 0.645	& 300&150 & 60 & --310 \\
			13	& 56649.2762& 0.649	& 300&130 & 35 & --349 \\
			14	& 56649.2856& 0.653	& 300&100 & 75 & --327 \\
			15	& 56649.2946& 0.657	& 300&90 & 104 & --318 \\
			16	& 56649.3027& 0.660	& 300&50 & 116 & --273 \\
			17	& 56649.3223& 0.669	& 300&70 & 159 & --253 \\
			18	& 56649.3303& 0.672	& 300&70 & 108 & --371 \\
			19	& 56649.3522& 0.682	& 300&90 & 128 & --304\\
			20	& 56649.3604& 0.685	& 300&100 & 130 & --353\\			
			\hline
		\end{tabular}
	\end{center}
\end{table*}

\section{Orbital Period Variation}

The best tool to analyze the orbital period variation is the O--C analysis. The available times of minima data of DN~Cas were collected from the literature and listed in Table~2 together with those determined in this study. The O--C residuals in Table~2 were calculated with the following light elements (Kreiner et al. 2001):

\begin{equation}
HJD_{\rm min}=~2441388.5662~+~2^{\rm d}.31095111~\times~E.
\end{equation}
where $HJD_{\rm min}$ and $E$ are the heliocentric time of conjunction for the primary minimum and the epoch number, respectively.

As listed in Table~2, the complete data set consists of 1 visual ($vis$), 21 photographic ($pg$), 7 photoelectric ($pe$) and 16 $ccd$ times of minima. The visual and photographic observations are relatively less accurate, therefore, the weighting scheme was chosen as follows: $vis$ and $pg=1$, $pe$ and $ccd=10$. 

The O--C residuals are plotted in Fig.~3. The relative accuracy of the $pe$ and $ccd$ data except in a few case is noteworthy. The residuals after the epoch number 2000 clearly shows a cyclic change which can be modeled by the light time effect (LTE) caused by gravitational disturbance of a distant companion to DN~Cas. Irwin (1959) derived the following equation (Eq.~2) to represent the O--C variations in the case of an existence of a distant companion:

\begin{equation}
\Delta t=\frac{a'_{12}\sin i'}{c}\left\lbrace\frac{1-e'^2}{1+e'\cos\nu'}sin(\nu'+\omega')+e'cos\omega'\right\rbrace 
\end{equation}
where $a'_{12}$, $i'$, $e'$ and $\omega'$ are the semi-major axis, orbital inclination, eccentricity and the argument of periastron of the long orbit (AB), respectively. Analyzing the times of minima data with Eq.~2 yielded the orbital period of the wide orbit as $P_{12}=42(9)$ years with an eccentricity of 0.37(0.2). Other orbital parameters of the distant companion are listed in Table~3. The best fitting LTE model is shown in Fig.~3.

\begin{table*}
	\small
	\begin{center}
		\caption{Times of minima of DN~Cas. \label{table2}}
		\begin{tabular}{cccccccc}
			\hline
			Times of minima & Method & Type & O--C & 			Times of minima & Method & Type & O--C \\
			HJD - 2400000   &        &      &      & 			HJD - 2400000   &        &      &      \\
			\hline
			28180.3100	& pg	&	sec	&	-0.0151	&	48172.3636	& pe	&	sec	&	0.0004	\\
			28427.6000	& pg	&	sec	&	0.0031	&	48619.5316	& pe	&	pri	&	-0.0006	\\
			28835.5100	& pg	&	pri	&	0.0302	&	49615.5498	& pe	&	pri	&	-0.0024	\\
			28865.4200	& pg	&	pri	&	-0.1021	&	50842.6600	& ccd	&	pri	&	-0.0072	\\
			29103.4800	& pg	&	pri	&	-0.0701	&	51107.2662	& pe	&	sec	&	-0.0049	\\
			29163.5900	& pg	&	pri	&	-0.0448	&	51422.7160	& ccd	&	pri	&	0.0001	\\
			29646.6500	& pg	&	pri	&	0.0264	&	51466.6180	& ccd	&	pri	&	-0.0060	\\
			29697.4800	& pg	&	pri	&	0.0155	&	52587.4405	& ccd 	&	pri	&	0.0052	\\
			29957.4700	& pg	&	sec	&	0.0235	&	53314.2254	& ccd	&	sec	&	-0.0040	\\
			30023.4000	& pg	&	pri	&	0.0914	&	53657.4062	& ccd	&	pri	&	0.0006	\\
			30321.3800	& pg	&	pri	&	-0.0413	&	53980.9378	& ccd	&	pri	&	-0.0010	\\
			30372.2600	& pg	&	pri	&	-0.0023	&	54050.2669	& ccd	&	pri	&	-0.0004	\\
			30582.5300	& pg	&	pri	&	-0.0288	&	54066.4437	& ccd	&	pri	&	-0.0003	\\
			30590.5500	& pg	&	sec	&	-0.0971	&	54838.3062	& ccd	&	pri	&	0.0045	\\
			39024.4840	& pg	&	pri	&	0.0208	&	55087.8900	& ccd	&	pri	&	0.0056	\\
			39053.3940	& pg	&	sec	&	0.0439	&	55872.4607	& ccd	&	sec	&	0.0084	\\
			39061.4030	& pg	&	pri	&	-0.0354	&	55879.3873	& pe 	&	sec	&	0.0022	\\
			41388.5710	& pg	&	pri	&	0.0048	&	55880.5433	& pe 	&	pri	&	0.0027	\\
			41602.3650	& pg	&	sec	&	0.0358	&	55903.6532	& ccd	&	pri	&	0.0031	\\
			44191.7650	& pg	&	pri	&	0.0151	&	55953.3388	& ccd	&	sec	&	0.0032	\\
			45044.4772	& pg	&	pri	&	-0.0137	&	55128.3279	& ccd	&	sec	&	0.0019	\\
			45653.4610	& vis	&	sec	&	0.0345	&	56287.2716	& ccd	&	pri	&	0.0036	\\
			47928.5576	& pe	&	pri	&	-0.0002	&		&	&		&		\\
			\hline
		\end{tabular}
	\end{center}
\end{table*}

\begin{table}
	\small
	\begin{center}
		\caption{Parameters of the wide orbit from the O--C analysis. \label{table3}}
		\begin{tabular}{lc}
			\hline
			Parameter                                & Value \\
			\hline
%			$T_0$ (HJD)                              & 2441388.5662(0.0002)\\
%			$P_{orb}$ (day)                          & 2.310951(0.000001)\\
%			$Q$ (day)                                & 0.0\\
			$T^{'}$ (HJD)                            & 2451700(280)\\
			$P_{12}$ (year)                          & 42(9)\\
			$a^{'}_{12}\sin i^{'}\,$(AU)             & 1.03(0.13)\\
			$e^{'}$                                  & 0.37(0.2)\\
			$\omega^{'}$(rad)                        & 5.0(0.5)\\
			$f(M_3)$ (M$_\odot$)                     & 0.0006(0.0002)\\
			%$M_3$ (M$_\odot$) for $i^{'}$=90$^\circ$ & \\
			%$M_3$ (M$_\odot$) for $i^{'}$=60$^\circ$ & \\
			%$M_3$ (M$_\odot$) for $i^{'}$=30$^\circ$ & \\
			\hline
		\end{tabular}
	\end{center}
\end{table}

\begin{figure*}
	\begin{center}
		\begin{tabular}{c}
			\hspace{1cm} \resizebox{100mm}{!}{\includegraphics{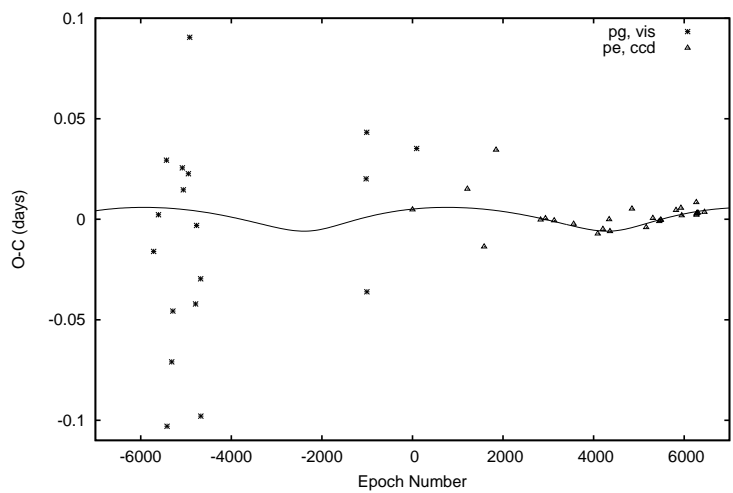}} \\
			\vspace{0.5cm}
		\end{tabular}
		\caption{O--C residuals and the best fitting LTE orbit.}\label{fig3}
	\end{center}
\end{figure*}

\section{Analysis of The Photometric and Radial Velocity Data}

The photometric data in the $UBVR_c$ and $I_c$-bands have been analyzed together with the radial velocity data of DN~Cas using the 2007 version of the Wilson-Devinney ({\sc wd}) code (Wilson \& Devinney, 1971). The initial values of the input parameters such as the orbital inclination, mass-ratio and relative radii of the components were adopted from the work of Hillwig et al. (2006). The orbital period and the conjunction time were adopted from the orbital period analysis (\S3) and fixed during the solution. The albedo and gravity darkening coefficients were set to 1.0 as suggested for early type stars. An infrared light source near DN~Cas with an angular distance of $5^{''}.3$ has been noted by Hillwig et al. (2006), which is in the area covered by our photometric aperture. Hence, the third light source has been considered and the third light contribution ($l_3$) has been converged during the solution. The orbital eccentricity was also taken into account and it was set as a free parameter during the first iterations.

The {\sc wd} code converged and found the best fitting model parameters after a few iterations. Initial system parameters were altered and the data were re-analyzed to check if the solution is stable and hence, final parameter set is reliable. Only a small value found for the eccentricity ($e=0.001$) was doubtful. However, its relatively big uncertainty $\Delta e=0.003$ decreased its reliability and allowed us to adopt a circular orbit although the system DN~Cas did not still complete its circularization time scale yet (see \S5.3). Nevertheless, it should be noted that the effect of such a small eccentricity on the determination of the physical parameters of the components is negligible.

In order to check the synchronicity of the components with the orbit, we modeled the spectral lines of the components and it is found that they rotate synchronously with the orbit (see Table~5). In the {\sc wd} code, this synchronous rotation is stated by fixing the \textit{F} parameter to 1.0 for each parameter to express the rotation rate of the component star.

The shapes of the components of close binary systems departs from sphere due to tidal effects causing a shift of the light centers from the mass center of the components. Due to the difference between the light centers and the mass centers of the component stars, the spectroscopic orbital model does not represent the true orbit (see the study by \"{O}zkarde\c{s} et al. (2009) on the highly distorted star CN~Hyi). In the case of DN Cas, both stars are well inside their roche lobes (see Table~4) and therefore nearly spherical. The final set of light curve and spectroscopic orbital parameters are listed in Table~4 and the model light curves together with the spectroscopic orbit are shown in Fig.~4. The system is detached and both components were found to be well inside their roche lobes.

The third light contribution was found to be negligible in all bands. As calculated by Hillwig et al. (2006), the light contribution of the third star is 14 per cent at $H$ \& $K$ infrared bands. This means that the light contribution of the third light source is not detectable at the $I$-band. The nature of the third light source is discussed in \S5.2.

\begin{table*}
\begin{center}
\small
\caption{Results from the simultaneous solution of light curves of DN~Cas. Adjusted and fixed parameters are presented in the separate panels of the table. Uncertainties of the adjusted parameters, as suggested by {\sc wd} code, are given in brackets.}
\label{table2}
\begin{tabular}{llc}\hline\hline
Parameter	&	Symbol              &  Value \\
\hline
Heliocentric primary conjunction	(HJD-240000)	&	$T_0$          & 56\,287.2699(0.0003) \\
Effective Temperature (pri)(K)	&	$T_{\rm eff1}$         &  \multicolumn{1}{c}{32100}\\
Effective Temperature (sec)(K)	&	$T_{\rm eff2}$         &    28020(100)   \\
Light contribution (pri) in U-band	&	$L_{1}/L_{1+2}$ ($U$)  &    0.666(0.006) \\
Light contribution (pri) in B-band	&	$L_{1}/L_{1+2}$ ($B$)  &    0.658(0.006) \\
Light contribution (pri) in V-band	&	$L_{1}/L_{1+2}$ ($V$)  &    0.657(0.006) \\
Light contribution (pri) in R$_c$-band&	$L_{1}/L_{1+2}$ ($R_c$)&    0.662(0.006) \\
Light contribution (pri) in I$_c$-band&	$L_{1}/L_{1+2}$ ($I_c$)&    0.658(0.006) \\
Orbital inclination ($^{\circ}$)&	$i$                    &    77.2(0.2)    \\
Mass ratio (M$_2$/M$_1$)	&	$q$            &  \multicolumn{1}{c}{0.722(0.002)}\\
Systemic velocity (km s$^{-1}$)	&$V_{\gamma}$          &    --45.5(0.3) \\
Surface potential (pri)	&	$\Omega_{\rm 1}$       &    4.03(0.01)    \\
Surface potential (sec)	&	$\Omega_{\rm 2}$       &    4.11(0.03)    \\
First critical Roche potential&	$\Omega_{\rm cr}$      &   \multicolumn{1}{c}{3.28}  \\
Relative radii	(pri)	&	$r_{\rm 1}$            &    0.307(0.002) \\
Relative radii	(sec)	&	$r_{\rm 2}$            &    0.246(0.002) \\
\hline
\multicolumn{3}{l}{Fixed parameters:}   \\
\hline
Orbital period (day)	&	$P$ (days)             & \multicolumn{1}{c}{2.310951}  \\
Orbital eccentricity	&	$e$                    &   0.0   \\
Rotation rate           &   $F$                    &   1.0   \\
			\hline
		\end{tabular}
	\end{center}
\end{table*}

\begin{figure*}
	\begin{center}
		\begin{tabular}{c}
			\hspace{6mm}\resizebox{93mm}{!}{\includegraphics{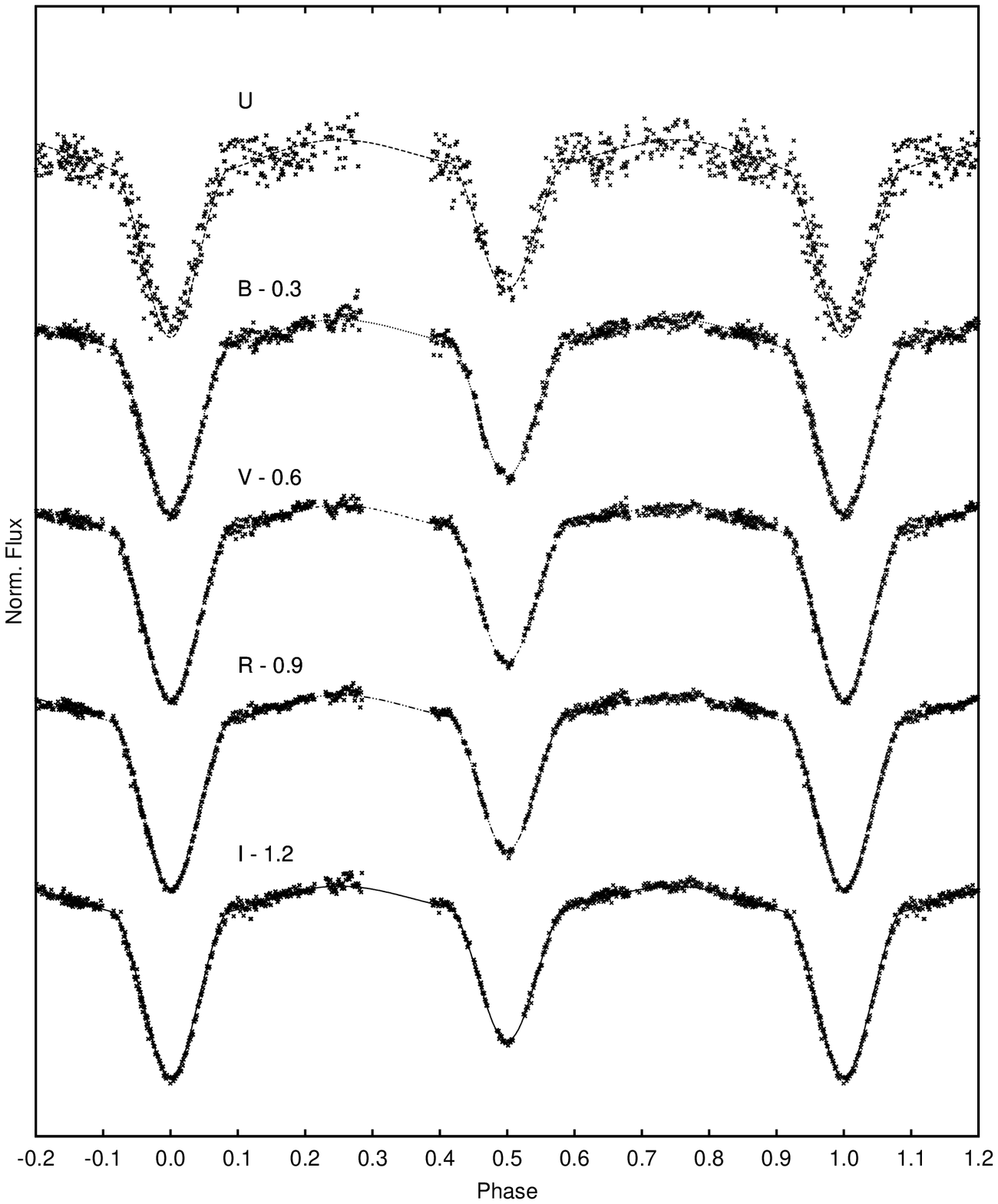}} \\
			\resizebox{100mm}{!}{\includegraphics{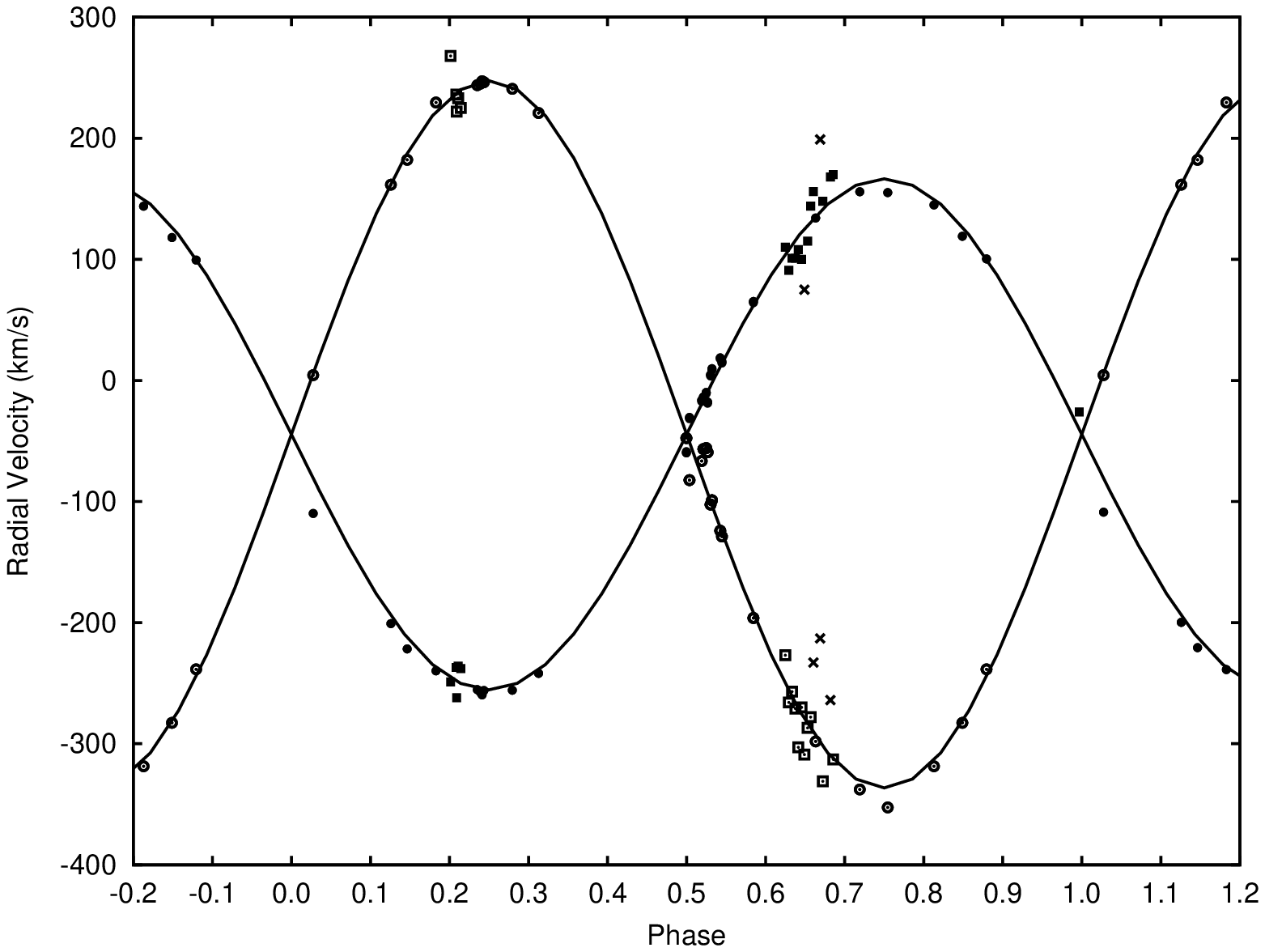}} \\
			\vspace{0.5cm}
		\end{tabular}
		\caption{Light curve (top) and spectroscopic orbit (bottom) models. The radial velocities shown by filled and empty circles are from Hillwig et al. (2006) and the filled and empty squares are obtained in this study. Radial velocities shown with crosses were not included in the analysis.}\label{fig4}
	\end{center}
\end{figure*}

\section{Discussion}

\subsection{Astrophysical parameters and the distance of the system}

The simultaneous solution of the light curve and the radial velocity data allowed us to obtain the physical parameters of the components of DN~Cas, which are presented in Table~5. The masses and radii of the components are derived with an uncertainty of $\sim$0.5 per cent. The masses of the components derived in this study is $\sim$1 per cent smaller than the masses derived by Hillwig et al. (2006), which is in the uncertainty box given for their mass determinations. Following the tables of stellar parameters given by Strai\v{z}ys \& Kurilene (1981), the physical parameters of the components imply that the spectral types of the primary and secondary components are B0V and B1V.

The distance of the system is calculated by deriving the unreddened color indices of the components with the $Q$-method of Johnson \& Morgan (1953). The unreddened color indices allowed us to derive the interstellar extinction of $A_v=3^{\rm m}.1(0.03)$. The absolute visual magnitude of the components has been derived by using the bolometric magnitude and the bolometric correction. Visual magnitude of the system, interstellar extinction and the absolute visual magnitude of the components has been used in the distance modulus to derive the distance of $1740\pm220$ pc for the system. 

The rotational velocities of the components were also measured by means of fitting synthetic spectra to the spectral lines of the components. The model atmosphere parameters were adopted from Table~5 and the suitable model atmospheres were selected from the Kurucz/Grids of model atmospheres\footnote{http://kurucz.harvard.edu/grids.html}. Modeling the spectral lines showed that both components rotate ($V_{rot1}=160\,$km s$^{-1}$, $V_{rot2}=130\,$km s$^{-1}$) in accordance with their synchronization rotational velocities given in Table~5.

\begin{table*}
	\setlength{\tabcolsep}{3pt} 
	\begin{center}
		\small
		\caption{Close binary stellar parameters of DN~Cas. Errors
			of parameters are given in brackets.} \label{table4}
		\begin{tabular}{lccc}\hline
			Parameter                 & Symbol  & Primary    & Secondary     \\
			\hline
			Spectral type             & Sp      & B0 V     & B1 V      \\
			Mass (M$_\odot$)          & \emph{M}& 19.04(0.07)  & 13.73(0.05) \\
			Radius (R$_\odot$)        & \emph{R}& 7.22(0.06)  & 5.79(0.06) \\
			Separation (R$_\odot$)    & \emph{a}& \multicolumn{2}{c}{23.53(0.05)} \\
			Orbital period (days)     & \emph{P} & \multicolumn{2}{c}{2.310951(0.000001)} \\
			Orbital inclination ($^{\circ}$) & \emph{i} & \multicolumn{2}{c}{77.2(0.2)} \\
			Mass ratio (M$_2$/M$_1$)    & \emph{q} & \multicolumn{2}{c}{0.722(0.002)} \\
			Eccentricity                 & \emph{e} & \multicolumn{2}{c}{0.0} \\
			Surface gravity (cgs)        & $\log g$ & 4.000(0.009)& 4.270(0.010) \\
			Colour index (mag)          & $B-V$ & \multicolumn{2}{c}{0.70(0.01)} \\
			Colour excess (mag)         &$E(B-V)$& \multicolumn{2}{c}{1.00(0.01)} \\
			Intrinsic colour index (mag)& $(B-V)_{\rm0}$ &\multicolumn{2}{c}{--0.30(0.01)} \\
			Visual absorption (mag)     & $A_{\rm v}$ &\multicolumn{2}{c}{3.10(0.03)} \\
			Component visual magnitude (mag)  & \emph{V}$_{1,2}$ & \multicolumn{1}{c}{7.29(0.03)} & \multicolumn{1}{c}{7.99(0.04)} \\
			Temperature (K)       & $T_{\rm eff}$  & 32100(1000) & 28020(1100) \\
			Luminosity (L$_\odot$)      & $\log$ \emph{L}& 4.70(0.06) & 4.27(0.08) \\
			Bolometric magnitude (mag)  &$M_{\rm bol}$   & --7.0(0.2) & --5.9(0.2) \\
			Abs. visual magnitude (mag)  &$M_{\rm v}$  & --3.9(0.2) & --3.2(0.2) \\
			Bolometric correction (mag)   &\emph{BC}& --3.1(0.1) & --2.7(0.1) \\
			Velocity amplitudes (km\,s$^{-1}$)&$K_{\rm 1,2}$& 211(3) & 292(5) \\
			Systemic velocity (km\,s$^{-1}$)  &$V_{\gamma}$ & \multicolumn{2}{c}{--45.5(0.3)} \\
			Observed rot. vel. (km\,s$^{-1}$)& V$_{rot}$   & 160(15) & 130(15) \\
			Computed sync. vel. (km\,s$^{-1}$)& V$_{synch}$   & 158(1) & 127(1) \\
			Distance (pc) &\emph{d} & \multicolumn{2}{c}{1740(220)} \\
			Proper Motions (mas yr$^{-1}$) & $\mu_{\alpha}\cos{\delta}$, $\mu_\delta$& \multicolumn{2}{c}{--2.57(1.60), --1.40(1.60)} \\
			Space Velocity Components (km\,s$^{-1}$) & \emph{U, W, V} & \multicolumn{2}{c}{43.0(9.6), --21.8(9.2), --18.1(13.4)} \\
			\hline
		\end{tabular}
	\end{center}
\end{table*}

\subsection{The unseen third companion}

The orbital period analysis of the system (\S3) yielded a cyclic period change, which was modeled with LTE orbit of $P_{12} =$ 42(9) yr. The mass function for this triple system is $f(m_3)=a'_{12}{\mathrm \sin}i'/P_{12}=0.0006\pm0.0002 M_\odot$. Assuming the inclination of the third component's orbit is $90^\circ$, the mass of the third star is found to be $0.88M_\odot$. In case, the orbit of the third star is co-planar ($i=i'=77^\circ.2$) with the close binary system, the mass of the third star is $0.90M_\odot$. If we assume a main-sequence star for the third companion, it should be a K0 type star and its light ratio in a system with B0 and B1 stars is much less than 1 per cent. Low light contribution of the third star can explain its invisibility in the visible region of the light curve and spectra.

\subsection{Evolutionary status of DN~Cas}

The evolutionary status of the components of DN~Cas have been investigated by locating the components in the HR-diagram together with the evolutionary tracks and isochrones of Bertelli et al. (2009) (see Fig.~5). The chemical composition of DN~Cas is not known, therefore, the isochrones in Fig.~5 have been calculated for two metallicity values ($Z=0.02$ and 0.04) in order to set the lower and upper limits of the age. Moreover, we have plotted the members of IC~1805 and IC~1848 open clusters in the same plane in Fig.~5 to see if there is an age difference. The best fitting isochrones yielded an age of $\tau=3$ Myr for $Z=0.04$ and $\tau=5$ Myr for $Z=0.019$ with and uncertainty of $\pm1$ Myr. Isochrones in Fig.~5 imply that the members of the two open clusters are 1-2~Myr younger than DN~Cas.

According to Zahn (1977) close binary systems tend to circularize their orbit due to tidal effects. Following his mathematical formulation, the circularization time scale for DN~Cas is $\sim$22$\pm$0.2 Myr. This circularization time scale is apparently bigger than the age of the system ($\tau=$3-5 Myr) but the difference is small compared to stellar evolution. This may imply a possible eccentricity for the orbit. However, the light curve and spectroscopic orbit analysis yielded a very small eccentricity ($e=0.001$) with a relatively big uncertainty box ($\Delta e=0.003$). Moreover, we do not know the effect of the third star on the orbit circularization. Therefore, higher quality observational data especially radial velocity measurements are needed to uncover the orbital eccentricity issue. One should note that a light curve alone does not provide a unique solution for $e$ and $\omega$ but $e{\mathrm cos}\omega$ for eclipsing binary systems (Kallrath \& Milone, 2009).

\begin{figure*}
\begin{center}
\begin{tabular}{c}
\resizebox{100mm}{!}{\includegraphics{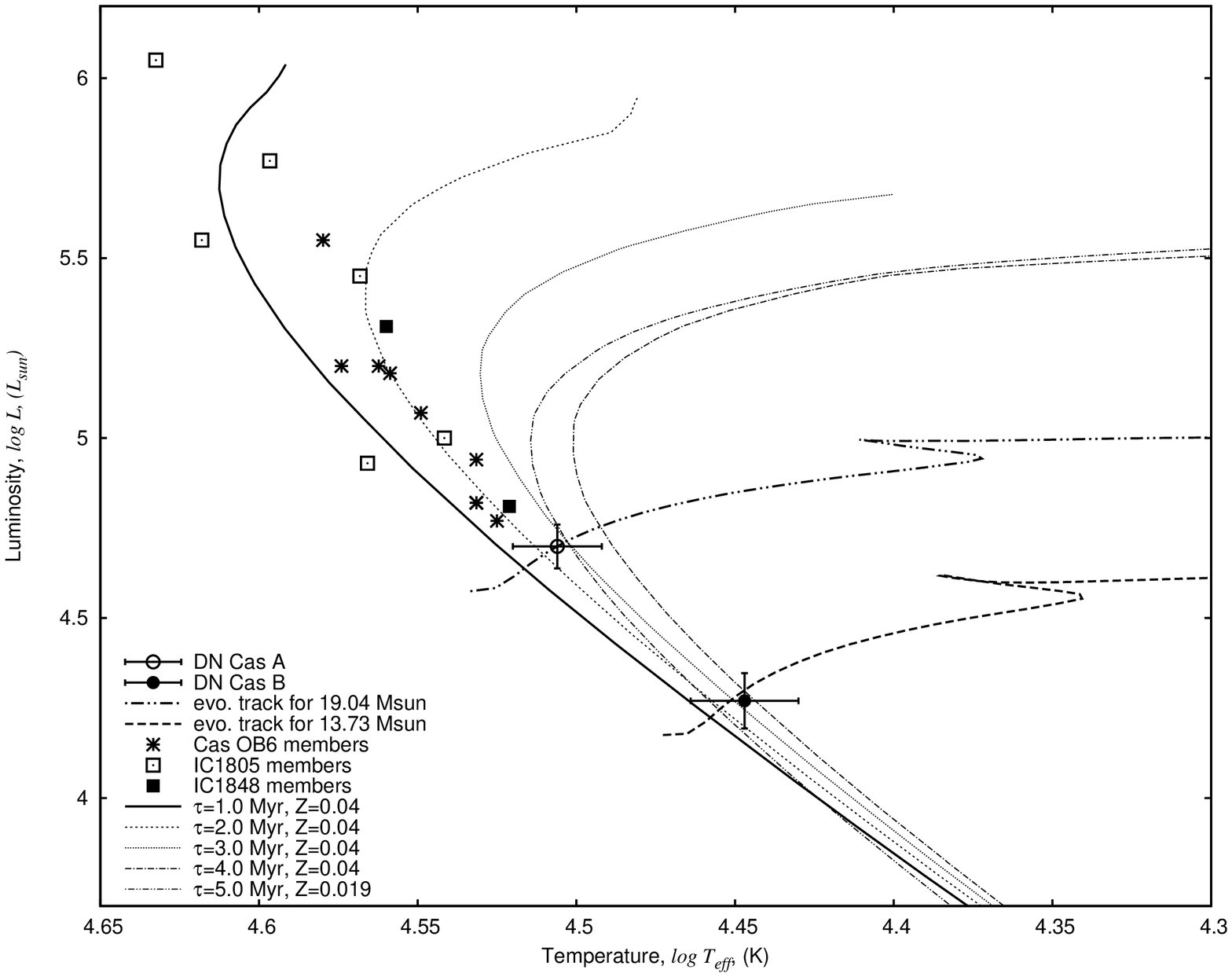}} \\
\vspace{0.5cm}
\end{tabular}
\caption{Location of DN~Cas components in the $\log~T_{eff}$ -- $\log~L$ plane together with the members of two nearby open clusters (IC~1805 and IC~1848).}\label{fig5}
\end{center}
\end{figure*}

\subsection{Kinematical Analysis and Possible Origin of DN~Cas}

To analyze the kinematical and dynamical properties of DN Cas, we have used the system's proper motion, the distance, and the velocity of the center-of-mass, which are given in Table~5. The proper motions were taken from the PPMXL Catalog of Positions and Proper Motions (Roeser et al. 2010), whereas the center-of-mass velocity and the distance are evaluated in this study. The system's space velocity with respect to Sun was calculated using Johnson \& Soderblom's (1987) algorithm where the space velocity components and their errors were found as ($U$, $V$, $W$)=(43.0$\pm$9.6, --21.8$\pm$9.2, --18.1$\pm$13.4) km s$^{-1}$ and given in Table~5. The first-order galactic differential rotation corrections were taken into account for precise space velocities, i.e. --34.1 and --2.9 km s$^{-1}$ for $U$ and $V$ space velocity components, respectively (Mihalas \& Binney 1981). The $W$ component is not affected from this first-order approximation. The velocity components are also corrected for the peculiar velocity of the Sun, which is 8.5$\pm$0.29, 13.38$\pm$0.43, 7$\pm$0.26 km s$^{-1}$ given by Co\c{s}kuno\u{g}lu et al. (2011). Thus, the final LSR space velocity components ($U$, $V$, $W$)$_{LSR}$ of DN~Cas are 17.4$\pm$9.6, --5.5$\pm$9.2, --11.6$\pm$13.4 km s$^{-1}$. So that, the total space velocity of DN~Cas is $S=21.6\pm18.9$ km s$^{-1}$. This value is in agreement with the space velocities of young thin-disc stars (Leggett 1992).

We also have performed test-particle integration in a Milky Way potentials of disc, halo, and bulge components. The galactic orbital parameters of the system were calculated within the integration time of 3 Gyr using the {\sc galpy} code of Bovy (2015). This integration time corresponds to about 15 revolutions around the Galactic centre, so that the orbital parameters could be determined reliably. The perigalactic ($R_{min}$) and apogalactic ($R_{max}$) distances of the system were found as 8.07 and 9.83 kpc, respectively. It turned out that the maximum distance of the system from the Galactic plane and the planer ellipticity of the system are $Z_{max}=203$ pc and about $e_{p}=0.10$, respectively. These values show that DN Cas system is orbiting the Galaxy almost in a circular orbit and that the system belongs to the galactic thin-disc population.

We have also compared the kinematic data of OB associations and open clusters in the vicinity of the binary system in order to investigate the possible birthplace of DN~Cas. There are two sources of kinematical data of associations, Mel'nik \& Dambis (2009) and Tetzlaff et al. (2010). The space velocity components given in these studies are almost the same. According to the galactic velocities, the DN~Cas is in agreement with Cas~OB1 and Cas~OB6 within the uncertainty box. However, the distance of Cas~OB6 is given as 1.8~kpc by Mel'nik \& Dambis (2009) and as 2.4~kpc by Tetzlaff et al. (2010), while the distance of Cas~OB1 is given as 2~kpc by Mel'nik \& Dambis (2010) and as 2~kpc by Tetzlaff et al. (2010). Nevertheless, there are distance determinations of the embedded clusters (see \S1) in Cas~OB6 which is in between 1.9-2.0~kpc. Moreover, according to the partition of Mel'nik \& Dambis (2009) (see Fig.~1), the direction of the binary goes through the Cas~OB6 borders and $\sim10\deg$ far from the Cas~OB1 
association. Consequently, the possibility of Cas~OB1 membership of DN~Cas is weaker than its OB6 membership. Considering a similar radial extension of the Cas OB6 association with its tangential extension (234 pc) given by Tetzlaff et al. (2010), the photometric distance of DN~Cas ($d=1.7\pm0.2$ kpc) implies that DN~Cas is within the borders of Cas~OB6. Moreover, the age of DN~Cas ($\tau=3-5$ Myr) which is 1-2 Myr older than the embedded open clusters in Cas OB6 is another supporting evidence for the Cas OB6 membership. Further evidence which would rely on accurately determined chemical composition is needed for a secure membership.

\section{Concluding Remarks}

The present study of DN~Cas, which is one of the massive members of Cas~OB6 association, could be concluded with the following remarks:

\begin{itemize}  
	\item The physical parameters of DN~Cas are obtained with high precision. These parameters allowed us to derive reliable distance and age for the system.
	
	\item Higher accuracy radial velocity measurements are necessary to investigate further if the very small eccentricity found in this study for the close orbit is true or spurious.
	
	\item The radial velocities given by Hillwig et al. (2010) perfectly matches the spectroscopic orbit without non-Keplerian effects. However, the model constructed for the system in the present study requires departures from pure-Keplerian motion in the radial velocities during the eclipses. More reliable radial velocities are necessary to clarify this discrepancy.
	
	\item It has been shown that DN~Cas has at least one physically bound distant companion with a minimum mass of 0.86 M$_\odot$ in a $P_{12}=42(9)$ yr orbit.
	
	\item DN~Cas is $\sim$1-2 Myr older than some known clusters (IC~1805 and IC~1848) in Cas~OB6 implying an age dispersion in stellar formation history of Cas~OB6.
\end{itemize}

\section{Acknowledgments}
The spectroscopic observations with TFOSC are granted by the T\"{U}B\.{I}TAK National Observatory with the project code 12ARTT150-255. This research has made use of ``Aladin sky atlas'' developed at CDS, Strasbourg Observatory, France.

\begin {thebibliography}{}
\harvarditem{Bakis}{2015}{B01} Bak{\i}\c{s} V., Hensberge H., Demircan O., Zejda M., Bilir S., Nitschelm C., 2015, ACPS, 496, 189.
\harvarditem{Bertelli}{2009}{B02} Bertelli G., Nasi E., Girardi L., Marigo P., 2009, A\&A, 508, 355.
\harvarditem{Bik}{2012}{B03} Bik A., Henning Th., Stolte A., Brandner W., Gouliermis D. A., Gennaro M., Pasquali A., Rochau B., Beuther H., Ageorges N., Seifert W., Wang Y., Kudryavtseva N., 2012, ApJ, 744, 13.
\harvarditem{Bonnarel}{2000}{B04} Bonnarel F., Fernique P., Bienaym\'{e} O., Egret D., Genova F., Louys M., Ochsenbein F., Wenger M., Bartlett J. G., 2000, A\&AS, 143, 33.
\harvarditem{Bovy}{2015}{B05} Bovy J., 2015, ApJS, 216, 29.
\harvarditem{Co{\c s}kuno{\v g}lu}{2011}{C01} Co{\c s}kuno{\v g}lu B., Ak S., Bilir S., Karaali S., Yaz E., Gilmore G., Seabroke G. M., Bienaym\'{e} O., Bland-Hawthorn J., Campbell R., Freeman K. C., Gibson B., Grebel E. K., Munari U., Navarro J. F., Parker Q. A., Siebert A., Siviero A., Steinmetz M., Watson F. G., Wyse R. F. G., Zwitter T., 2011, MNRAS, 412, 1237.
\harvarditem{Frazier\&Hall}{1974}{F01} Frazier T. H., Hall D. S., 1974, PASP, 86, 661.
\harvarditem{Hillwig}{2006}{H01} Hillwig Todd C., Gies Douglas R., Bagnuolo William G. Jr., Huang Wenjin, McSwain M. Virginia, Wingert David W., 2006, ApJ, 639,1069.
\harvarditem{Hoffmeister}{1947}{H02} Hoffmeister C., 1947, Berlin : Akademie-Verlag, 41.
\harvarditem{Irwin}{1959}{I01} Irwin John B., 1959, AJ, 64, 149.
\harvarditem{Johnson\&Morgan}{1953}{J01} Johnson H. L., Morgan W. W., 1953, ApJ, 117, 313.
\harvarditem{Johnson\&Soderblom}{1987}{J02} Johnson D. R. H., Soderblom D. R., 1987, AJ, 93, 864.
\harvarditem{Kallrath\&Milone}{2009}{K01} Kallrath J., Milone E. F., 2009, Eclipsing Binary Stars: Modeling and Analysis, Springer-Verlag New York.
\harvarditem{Kopal}{1942}{K02} Kopal, Z. 1942, Proc. U.S. Natl. Acad. Sci., 28, 133.
\harvarditem{Kopal}{1945}{K03} Kopal, Z. 1945, Proc. Am. Phil. Soc. 89, 517.
\harvarditem{Kreiner}{2001}{K02} Kreiner Jerzy M., Kim Chun-Hwey, Nha Il-Seong, 2001, An Atlas of O-C Diagrams of Eclipsing Binary Stars, Cracow, Poland: Wydawnictwo Naukowe Akademii Pedagogicznej.
\harvarditem{Leggett}{1992}{L01} Leggett S. K., 1992, ApJS, 82, 351.
\harvarditem{Lim}{2014}{L02} Lim Beomdu, Sung Hwankyung, Kim Jinyoung S., Bessell Michael S., Karimov R., 2014, MNRAS, 438, 1451.
\harvarditem{Mason}{1998}{M01} Mason B. D., Gies D. R., Hartkopf W. I., Bagnuolo W. G. Jr., ten Brummelaar T., McAlister H. A., 1998, AJ, 115, 821. 
\harvarditem{Melnik\&Dambis}{2009}{M02} Meln'ik A., M., Dambis, A. K., 2009, MNRAS, 400, 518.
\harvarditem{Mihalas\&Binney}{1981}{M03} Mihalas D., Binney J., 1981, Galactic astronomy: Structure and kinematics (2nd edition), San Francisco, CA, W. H. Freeman and Co., 1981. 608 p.
\harvarditem{Ozkardes}{2009}{O01} \"{O}zkarde\c{s} B., Erdem A., Bak{\i}\c{s} V., 2009, NewA, 14, 461.
\harvarditem{Roeser}{2010}{R01} Roeser S., Demleitner M., Schilbach E., 2010, AJ, 139, 2440.
\harvarditem{Straizys}{1981}{S01} Strai\v{z}ys V., Kuriliene G., 1981, Ap\&SS, 80, 353.
\harvarditem{Straizys}{2013}{S02} Strai\v{z}ys V., Boyle R. P., Janusz R., Laugalys V., Kazlauskas A., 2013, A\&A, 554, 3.
\harvarditem{Tetzlaff}{2010}{T01} Tetzlaff N., Neuh{\"a}user R., Hohle M.~M., Maciejewski G., 2010, MNRAS, 402, 2369.
\harvarditem{Wilson\&Devinney}{1971}{W01} Wilson R. E., Devinney E. J., 1971, ApJ, 166, 605.
\harvarditem{Zahn}{1977}{Z01} Zahn J. P., 1977, A\&A, 57, 383.
\harvarditem{Zakirov}{2001}{Z02} Zakirov M. M., 2001, KFTN, 17, 313.
\end{thebibliography}

\end{document}